\documentclass[12pt]{iopart}
\usepackage{iopams} 
\usepackage[usenames,dvipsnames]{xcolor} 
\begin{document}

\title[Stability of event horizons against neutrino flux]{Stability of event horizons against neutrino flux: The classical picture}

\author{Koray D\"{u}zta\c{s}}

\address{Bo\u{g}azi\c{c}i University, Department of Physics \\ Bebek 34342, \.Istanbul, Turkey}
\ead{koray.duztas@boun.edu.tr;duztasko@hotmail.com}
\vspace{10pt}

\begin{abstract}
It has been claimed that cosmic censorship conjecture can be violated by quantum tunnelling of neutrinos, though it is strongly supported by classical arguments. We  consider the classical interaction of an extremal Kerr black hole with a test massless Dirac field, i.e. a ``neutrino field". Evaluating the flux integrals imposed by the energy momentum tensor for fermionic fields and the Killing vectors of the space-time, we prove that this interaction can indeed destroy the event horizon of the black hole and convert it to a naked singularity. The classical process leads to a more generic violation of cosmic censorship conjecture compared to quantum tunnelling processes which occur with a low probability. The range of frequencies of the test neutrino field that can be used to destroy the black hole turns out to be the superradiant range for bosonic fields. We comment on back reaction and quantum effects. We argue that the destruction of black holes by neutrino fields  cannot be fixed by self-force effects unlike similar attempts involving test bodies and bosonic fields.
\end{abstract}

\pacs{04.20.Dw, 04.20.Gz, 04.70.Bw}
%
\vspace{2pc}
\noindent{\it Keywords}: cosmic censorship, neutrino fields, spinor formalism
%
\submitto{\CQG}
%
%
%

\section{Introduction}

A space-time possesses  a singularity if it fails to satisfy causal geodesic completeness which requires that every time-like and null geodesic can be extended to arbitrarily large affine parameter value both into the future and into the past. In essence, causal geodesic completeness means that photons or freely moving particles can not appear from or disappear off the edge of the universe.

According to the singularity theorems of Hawking and Penrose  \cite{singtheo}, a space-time $M$  can not satisfy causal geodesic completeness if --in addition to some generic conditions-- Einstein's equations are satisfied and $M$ contains a \emph{trapped surface}. Trapped surfaces arise when the gravitational collapse of a localized body (e.g. a star) to within its Schwarzschild radius takes place, provided that the deviation of the model from spherical symmetry is negligible. Thus, a singularity ensues as a result of gravitational collapse in classical general relativity.

If a space-time contains a well defined external  infinity ${\cal I}^+$ the black hole region  of the space-time is defined by (see e.g. \cite{waldbook})
\begin{equation}
\mathcal{B}=M-J^-(\mathcal{I}^+)
\end{equation}
with an \emph{event horizon} that constitutes the boundary of $\mathcal{B}$ in $M$.
\begin{equation}
\mathcal{H}=\partial \mathcal{B}=\dot{J}^-(\mathcal{I}^+)\cap M
\end{equation}
where $J^-$ denotes  the causal past of a set and $\dot{J}$ denotes its boundary. In an asymptotically predictable space-time $M$, a trapped surface $\mathcal{T}$ is entirely contained within the \emph{black hole} region $\mathcal{B}$; i.e. $\mathcal{T}\subset \mathcal{B}$. Therefore if the singularity has formed in the way prescribed by Hawking and Penrose \emph{it is hidden behind an event horizon}, so it is invisible to the rest of the space-time. In the opposite case  where the space-time contains a singularity that lies to the past of ${\cal I}^+$, so time-like curves may be drawn into the past that terminate on the singularity, the singularity is said to be \emph{naked}. Whether a singularity is \emph{naked} or \emph{clothed} is crucial to preserve causal structure. In the presence of naked singularities, initial conditions on a Cauchy surface become undefined since the surface necessarily intersects the singularity, thus asymptotic predictability is disabled. Causal behaviour is characterized by prohibiting the existence of closed time-like or causal curves. It turns out that for asymptotically flat space-times, closed time-like curves which violate causality can evolve in the regions which contain naked singularities \cite{tipler}. It becomes impossible to predict the behaviour of space-time in the causal future of a singularity.

It is an open question whether physically realistic collapse situations may arise without trapped surfaces, resulting in singularities that are not necessarily hidden behind an event horizon. For example the singularity theorem does not apply in the presence of a positive cosmological constant so it is unknown how the collapse will develop in that case. This led  Penrose to propose the \emph{Cosmic Censorship Conjecture}  (CCC) in 1969 \cite{penrose.orig.ccc}, to avoid the physical and philosophical pathologies that could arise in a space-time containing a naked singularity. CCC asserts that gravitational collapse of a body always ends up in a black hole rather than a naked singularity,
so all singularities are hidden within the event horizons of black holes; i.e. they are invisible to distant observers (see \cite{waldrev} for a review). Distant observers neither encounter any singularities nor any effects propagating out of singularities. Conjecturing singularities to be hidden behind event horizons without any access to distant observers allows us to preserve causal structure despite the fact that  the formation of singularities is inevitable in classical general relativity.

As a concrete proof of CCC has been elusive, it has become customary to attack the closely related --though not identical-- problem of the stability of event horizons in the interactions of the black hole with test particles or fields. In these problems the initial state is a stationary Kerr-Newman space-time uniquely defined by three parameters (Mass $M$, charge $Q$, and angular momentum per unit mass $a$), satisfying
\begin{equation}
M^{2} \geq Q^{2}+a^{2}. \label{criterion}
\end{equation}
which defines a black hole surrounded by an event horizon as opposed to a naked singularity. To test the stability of the event horizon one lets the black hole absorb some particles or fields coming from infinity. At the end of the interaction the space-time is expected to settle to another stationary configuration with new values of $M$, $Q$, and $a$. Then one may check if it is possible to increase charge or angular momentum of the black hole beyond the extremal limit saturating (\ref{criterion}) so that the final configuration of the parameters violates (\ref{criterion}) and defines a naked singularity.

The first thought experiment to test the stability of black hole horizons was constructed by Wald \cite{wald74}. He showed that particles having enough charge or angular momentum to exceed extremality are not captured by the black hole. This work motivated many authors to construct similar thought experiments, including some recent attempts involving neutrinos. Richartz and Saa argued that, though classical results provide a strong support for the validity of CCC, it is possible to over-spin a nearly extremal Kerr black hole \cite{ri_saa_1}, and over-charge a nearly extremal Reissner-Nordst\"{o}m black hole \cite{ri_saa_2} by means of quantum tunneling of neutrinos. In the present work we consider the \emph{classical} scattering of \emph{neutrino fields} from an extremal Kerr black hole saturating (\ref{criterion}). We show that the classical process leads to an even more generic violation of CCC compared to quantum tunnelling processes which occur with a low probability. 

We follow the recipes developed in \cite{dkn} and \cite{duztasem}, which evaluate the bosonic cases of spin-0 (scalar field) and spin-1 (electromagnetic field) respectively. Neutrino field refers to a test massless Dirac field in Kerr background. This Kerr background, uniquely parametrized by $M$ and $a$, is the initial stationary state of the problem. The impact of the test field on the background geometry is negligible. The field coming  in from infinity, is partially transmitted through the horizon and partially reflected back to infinity.  To calculate the changes in the black hole parameters, we construct expressions for the fluxes of energy and angular momentum carried by the neutrino field and evaluate them at the horizon of the black hole.  We treat the free neutrino test field using Newman-Penrose (NP) \cite{newpen} formalism. We use the separation of Dirac equations for the relevant NP variables in Kerr space-time by Chandrasekhar \cite{chandradirac}, and asymptotic solutions at the horizon and at infinity for the massless case by Teukolsky \cite{teuk2}. In this context we check if neutrino fields can be used to over-spin an extremal Kerr black hole, in the context of classical general relativity. 

\section{Neutrino fields in Kerr geometry}
\label{sec:1}
This section consists of a brief review of spinor formalism and previous results involving neutrino fields in Kerr space-time especially by Chandrasekhar and Teukolsky. We start with Dirac equation which couples two fermion fields via
\begin{eqnarray}
& &\nabla_{A A^{\prime}}P^{A} + \rmi \mu_f \bar{Q}_{A^{\prime}}=0 \nonumber \\
& & \nabla_{A A^{\prime}}Q^{A} + \rmi \mu_f \bar{P}_{A^{\prime}}=0   \label{diraceqn} 
\end{eqnarray}
where $\nabla_{A A^{\prime}}$ is the spinor covariant derivative defined axiomatically as a map $\nabla_x=\nabla_{X X^{\prime}}: \theta_{...} \mapsto \theta_{...;X X^{\prime}}$ \cite{agr} and $\sqrt{2}\mu_f$ is the mass of the fermion field. 

Every spin basis induces a tetrad of null vectors.
\begin{equation}
l^a=o^A \bar{o}^{A'} \quad n^a=\iota^A \bar{\iota}^{A'} \quad m^a=o^A \bar{\iota}^{A'} \quad \bar{m}^a=\iota^A \bar{o}^{A'} \label{nptetrad}
\end{equation}
$l$ and $n$ are real while $m$ and $\bar{m}$ are complex (conjugates). Null tetrad satisfies orthogonality relations
\begin{eqnarray}
& &l_an^a=n_al^a=-m_a \bar{m}^a=-\bar{m}_a m^a=1 \nonumber \\
& &l_am^a=l_a \bar{m}^a=n_am^a=n_a\bar{m}^a=0
\end{eqnarray}
In 1962, Newman and Penrose \cite{newpen} have developed the idea of adapting tetrads of null vectors to spinors. The directional derivatives along the null directions are denoted by conventional symbols
\begin{equation}
D=l^a\nabla_a,\quad \Delta=n^a\nabla_a,\quad \delta=m^a\nabla_a,\quad \bar{\delta}=\bar{m}^a\nabla_a
\end{equation}
$\nabla_a$ can be expressed as a linear combination of these operators. 
\begin{eqnarray}
\nabla_a &=&g_a^{\;\;b}\; \nabla_b \nonumber \\
&=& (n_al^b+l_an^b-\bar{m}_am^b-m_a\bar{m}^b)\nabla_b  \nonumber \\
&=& n_aD + l_a \Delta - \bar{m}_a \delta -m_a \bar{\delta} \label{nablanp}
\end{eqnarray}
Dirac's equations (\ref{diraceqn}) can explicitly be written in the form:
\begin{eqnarray}
& &(D+\epsilon-\rho)P^0+ (\bar{\delta}+\pi -\alpha)P^1=\rmi\mu_f \bar{Q}^{\dot{1}} \label{dirac1} \\
& &(\Delta+\mu -\gamma )P^1 + (\delta -\tau +\beta )P^0=-\rmi\mu_f \bar{Q}^{\dot{0}} \label{dirac2} \\
& & (D+ \bar{\epsilon} -\bar{\rho} )\bar{Q}^{\dot{0}}+ (\delta +\bar{\pi} -\bar{\alpha})\bar{Q}^{\dot{1}}=-\rmi\mu_f P^1 \label{dirac3} \\
& & (\Delta +\bar{\mu} - \bar{\gamma})\bar{Q}^{\dot{1}} +(\bar{\delta} + \bar{\beta} - \bar{\tau})\bar{Q}^{\dot{0}}=\rmi \mu_f P^0 \label{dirac4}
\end{eqnarray}
where $P^0$,$Q^0$ and $P^1$,$Q^1$ are components of $P^A$,$Q^A$ along the spinor dyad basis $o^A$ and $\iota^A$ respectively. Chandrasekhar \cite{chandradirac} showed that equations (\ref{diraceqn}) can be solved by a separation of variables.
\begin{eqnarray}
P^0 &=& (r-ia\cos \theta)^{-1}[{}_{-1/2}R(r)][{}_{-1/2}S(\theta)] \rme^{-\rmi\omega t }\rme^{\rmi m\phi} \nonumber \\
P^1 &=& [{}_{1/2}R(r)][{}_{1/2}S(\theta)] \rme^{-\rmi\omega t }\rme^{\rmi m\phi} \nonumber \\
Q^{\dot{0}} &=&  -(r+ia\cos \theta)^{-1}[{}_{-1/2}R(r)][{}_{1/2}S(\theta)] \rme^{-\rmi\omega t }\rme^{\rmi m\phi} \nonumber \\
Q^{\dot{1}} &=&  {[}_{1/2}R(r)][{}_{-1/2}S(\theta)] \rme^{-\rmi\omega t }\rme^{\rmi m\phi}
\label{diracsep}
\end{eqnarray}
This separation leads to a pair of equations for both $[{}_{\pm1/2}R(r)]$ and $ [{}_{\pm1/2}S(\theta)]$ which can be expressed as a single equation with $s=\pm 1/2$ as a parameter. As we let $\mu_f =0$ for neutrino fields the radial equation takes the form

\begin{equation}
\Delta^{-s} \frac{\partial}{\partial r} \left(\Delta^{s+1} \frac{\partial ({}_sR)}{\partial r} \right) + \left( \frac{K^2 -2\rmi s(r-M)K}{\Delta} +4\rmi s \omega r -\lambda \right)({}_sR)=0 \label{teukradial}
\end{equation}

where $K \equiv (r^2+a^2)\omega - am$ and $\lambda \equiv A+ a^2 \omega^2 -2am\omega$. The angular equation is given by

\begin{eqnarray}
&&\frac{1}{\sin \theta}\frac{\partial}{\partial \theta} \left( \sin \theta \frac{\partial S}{\partial \theta} \right) \nonumber \\
&& +\left( a^2\omega^2 \cos^2 \theta -\frac{m^2}{\sin^2 \theta}-2a\omega s \cos \theta -\frac{2ms\cos \theta}{\sin^2 \theta}-s^2 \cot^2 \theta +s+A \right)S=0 \nonumber \\
&& \label{teukangular}
\end{eqnarray}

These are exactly Teukolsky's equations for massless fields. The asymptotic solutions at infinity for the radial functions are \cite{teuk2}
\begin{eqnarray}
& &\left. Y_{\rm{in}}  \frac{\rme^{-\rmi \omega r*}}{r} \quad , \quad Y_{\rm{out}}  \frac{\rme^{\rmi \omega r*}}{r^2} \right \} s=1/2 \nonumber  \\
& &  \left. Z_{\rm{in}}  \frac{\rme^{-\rmi \omega r*}}{r} \quad , \quad Z_{\rm{out}}  \rme^{\rmi \omega r*} \right \} s=-1/2 \label{teuksolnsinf}
\end{eqnarray}
where we have adopted the notation of Teukolsky and Press \cite{teukII,teukIII}; $Y_{\rm{in}}$, $Y_{\rm{out}}$, $Z_{\rm{in}}$, $Z_{\rm{out}}$ are the normalizations of the ingoing and outgoing waves at infinity for the cases $s=1/2$ and $s=-1/2$ respectively, and $r^*$ is the tortoise coordinate defined by $dr^*/dr=(r^2+a^2)/\Delta$, so that $r^* \to -\infty$ as the horizon is approached. 

Only the ingoing solutions are physical at the horizon \cite{teukIII}. Therefore the  asymptotic solutions of the radial equation (\ref{teukradial}) near the horizon are given by
\begin{eqnarray}
 && \left. Y_{\rm{hole}}(\Delta^{-1/2}) \rme^{-\rmi kr*} \right\} s=1/2 \nonumber \\
 & &\left.  Z_{\rm{hole}}(\Delta^{1/2})  \rme^{-\rmi kr*}  \right\} s=-1/2
\label{teuksolnshor}
\end{eqnarray}
where $k=\omega-m\Omega$, $\Omega=a/2Mr_+$ is the rotational frequency of the black hole.
 
The angular equation (\ref{teukangular}) constitutes a Sturm-Liouville eigenvalue problem for the separation constant $A={}_sA^m_{\;\; l}(a \omega)$, together with boundary conditions of regularity at $\theta=0$ and $\pi$. From Sturm-Liouville theory the eigenfunctions ${}_sS^m_{\;\; l}$ are complete and orthogonal on $0 \leq \theta \leq \pi$ for each $s$ and $a\omega$. One can also define  ${}_sZ^m_{\;\; l}= {}_sS^m_{\;\; l}\rme^{\rmi m\phi}$ with orthonormalization:

\begin{equation}
\int_0^{2\pi} d \phi\int_0^{\pi} d \theta({}_s Z_{lm}(\theta,\phi, a\omega)) ({}_s Z^*_{l'm'}(\theta,\phi, a\omega)) \sin \theta =\delta_{ll'}\delta_{mm'} \label{anguortho}
\end{equation}
The functions ${}_s Z_{lm}(\theta,\phi a\omega)$ reduce to spin weighted spherical harmonics ${}_s Y_{lm}(\theta,\phi)$ when $a\omega=0$.

\section{The stability of the event horizon}
In this section we evaluate the changes in the mass and angular momentum of the black hole due to its interaction with the test neutrino field. For that purpose we are going to use the current conservation equations imposed  by the Killing vectors of the space-time and the local conservation of  energy and momentum in general relativity. If $K$ is a Killing vector, by definition it satisfies  $\mathcal{L}_K g=0$, where $\mathcal{L}$ is the Lie derivative. This can be re-arranged to give the Killing equation $\nabla_{(a}K_{b)}=0$. The current conservation equation $\nabla_a (T^{ac}K_c)=0$ is derived by combining the expression  of the space-time symmetry in terms of Lie derivatives or equivalently the Killing equation with the local conservation of energy-momentum $\nabla_c T^{ac}=0$, where $T_{ac}$ is the energy momentum tensor. This allows us to express the rates of change in the corresponding black hole parameters as fluxes into the black hole. Kerr spacetime is stationary and axi-symmetric with corresponding  Killing vectors $\partial /\partial t$ and $\partial / \partial \phi$. Therefore the net radial flux of energy and the net radial flux of angular momentum across any sphere centered at the black hole are given by surface integrals of $-T^{r}_{\;\;t}$ and $T^{r}_{\;\;\phi}$ respectively.
\begin{equation}
\left(\frac{dM}{dt}\right)_{\rm b.h} = - \int_{S} \sqrt{-g} \, T^{1}_{\;\;0} d\theta d\phi
                                                   \label{eq:dm/dt}
\end{equation}
Since $dM=dE$ for the black hole, and
\begin{equation}
\left(\frac{dL}{dt}\right)_{\rm b.h}
=   \int_{S} \sqrt{-g} \, T^1_{\;\;3} d\theta d\phi \label{eq:dl/dt}
\end{equation}
where the label b.h. stands for black hole..

To test the stability of the horizon of a Kerr black hole, we define an indicator
\begin{equation}
{\mathcal{C}} =  M^{2} - a^{2}
\end{equation}
Then, using $a=L/M$
\begin{equation}
\delta {\mathcal{C}}  =   \int \frac{d{\mathcal{C}}}{dt}  dt  =  \int \frac{2}{M} \left\{ (M^{2}+a^{2}) \frac{dM}{dt} - a \frac{dL}{dt} \right\} dt \label{deltaccc}
\end{equation} 
implying
\begin{equation}
 \frac{d{\mathcal{C}}}{dt} = \int_S \sqrt{-g} [ (M^{2}+a^{2})(- T^{1}_{\;\;0})   - a T^{1}_{\;\;3} ] d \theta \; d \phi \label{eq:dC/dt1}
\end{equation}
For an extremal black hole saturating the main criterion (\ref{criterion}) $\delta {\mathcal{C}}$ should always remain positive to preserve the event horizon. If the initial state is an extremal black hole and $\delta {\mathcal{C}}$ has a negative value, the final state describes a naked singularity.
 
Kerr space-time can be represented by an NP tetrad of the form:
\begin{eqnarray}
l^\mu & = & [(r^2 + a^2)/\Delta ,1,0,a/\Delta],\nonumber \\
n^\mu & = & [(r^2+a^2), -\Delta ,0,a]/(2\Sigma) \nonumber \\
m^\mu & =& [ia \sin \theta ,0,1, i/\sin \theta]/[\sqrt{2}(r+ia \cos \theta)]
\label{tetrad0}
\end{eqnarray}
where $\Sigma=r^2+a^2\cos^2\theta$ and $\Delta=r^2-2Mr+a^2$, which should not be confused with the NP derivative operator $\Delta$. Note that $\Delta \to r^2-2Mr+a^2+Q^2$ gives the Kerr-Newman space-time, and $a\to 0$ gives the Schwarzschild space-time. Using tetrad (\ref{tetrad0}) one can derive
\begin{equation}
T_{ab}l^al^b \frac{\Delta^2}{4\Sigma}-\Sigma T_{ab}n^an^b= -(r^2+a^2) T^{1}_{\;\;0}- T^{1}_{\;\;3}
\label{lalbnanb0}
\end{equation}
We are going to evaluate the fluxes (\ref{eq:dm/dt}) and (\ref{eq:dl/dt}) at the horizons ($r=r_+$) of extremal Kerr black holes ($r_+=M$). In that case  we recognize that the right-hand-side of (\ref{lalbnanb0}) is the integrand of (\ref{eq:dC/dt1}). 

For fermionic fields, the energy momentum tensor in terms of the corresponding NP scalars is given by (see \cite{guven}, \cite{scatter})

\begin{eqnarray}
T_{AA'BB'}&=&-\frac{1}{2}\rmi \{ P_A \nabla_{BB'}\bar{P}_{A'}- \bar{P}_{A'}\nabla_{BB'}P_A + P_B \nabla_{AA'}\bar{P}_{B'} -\bar{P}_{B'}\nabla_{AA'}P_B \nonumber \\
&-& Q_A \nabla_{BB'}\bar{Q}_{A'}+ \bar{Q}_{A'}\nabla_{BB'}P
Q_A - Q_B \nabla_{AA'}\bar{Q}_{B'} +\bar{Q}_{B'}\nabla_{AA'}Q_B \}\nonumber \\
& & 
\label{tmunu}
\end{eqnarray} 
implying
\begin{eqnarray}
T_{ab}l^al^b&=&T_{AA'BB'}o^A \bar{o}^{A'} o^B\bar{o}^{B'} \nonumber  \\
&=& -\rmi\{P^1 D \bar{P}^{\dot{1}} - \bar{P}^{\dot{1}} DP^1 + \bar{Q}^{\dot{1}} DQ^1-Q^1 D \bar{Q}^{\dot{1}} \} \nonumber  \\
& & \label{Tablalb}
\end{eqnarray}
and
\begin{eqnarray}
T_{ab}n^an^b&=&T_{AA'BB'}\iota^A \bar{\iota}^{A'} \iota^B\bar{\iota}^{B'} \nonumber  \\
&=& -\rmi \{P^0 \Delta \bar{P}^{\dot{0}} - \bar{P}^{\dot{0}} \Delta P^0 + \bar{Q}^{\dot{0}}\Delta Q^0-Q^0 \Delta \bar{Q}^{\dot{0}} \} \nonumber  \\
& &\label{Tabnanb}
\end{eqnarray}
where $D$ and $\Delta$ are NP derivative operators. Now we may plug in the solutions near the horizon for $P^0$ and $P^1$ following (\ref{diracsep}) and (\ref{teuksolnshor}) to evaluate the left-hand-side  of (\ref{lalbnanb0}).  We first write the general solution in terms of separated modes:

\begin{eqnarray}
& &P^1= \int d\omega \rme^{-\rmi\omega t} \sum_{l,m} \rme^{\rmi m\phi} (S_+) Y_{\rm{hole}}(\Delta^{-1/2}) \rme^{-\rmi kr*} \nonumber \\
& &\bar{Q}^{\dot{1}}= \int d\omega \rme^{-\rmi\omega t} \sum_{l,m} \rme^{\rmi m\phi} (S_-) Y_{\rm{hole}}(\Delta^{-1/2}) \rme^{-\rm ikr*} \nonumber \\
& &P^0= \int d\omega \rme^{-\rmi\omega t} \sum_{l,m} \rme^{\rmi m\phi} (S_-)(-\rho) Z_{\rm{hole}}(\Delta^{1/2}) \rme^{-\rmi kr*} \nonumber  \\
& &\bar{Q}^{\dot{0}}= \int d\omega \rme^{-\rmi\omega t} \sum_{l,m} \rme^{\rmi m\phi} (S_+)(\rho^*) Z_{\rm{hole}}(\Delta^{1/2}) \rme^{-\rmi kr*} \nonumber  \\
& &\label{p1q1p0q0}
\end{eqnarray}
where $\rho=-(r-ia\cos \theta)^{-1}$, and the dependence of $Y_{\rm{hole}}$, $Z_{\rm{hole}}$ on $l,m,\omega$ are implied. Then

\begin{eqnarray}
P^1 D \bar{P}^{\dot{1}}&=&\Delta^{-2}\int d\omega d \omega' \sum \vert Y \vert^{2} \rme^{-\rmi(k-k')r*} (S_+S_+') \rme^{\rmi(m-m')\phi} \rme^{-\rmi(\omega- \omega') t} \nonumber \\
& & \times [\rmi(r^2 + a^2)(w'+k') - (r-M) - \rmi m'a] \label{term1}
\end{eqnarray}

\begin{eqnarray}
\bar{P}^{\dot{1}} D P^1&=&\Delta^{-2}\int d\omega d \omega' \sum \vert Y \vert^{2} e^{-i(k-k')r*} (S_+S_+') e^{i(m-m')\phi} e^{-i(\omega- \omega') t} \nonumber \\
& & \times [-i(r^2 + a^2)(w+k) - (r-M) + ima]
\end{eqnarray}

\begin{eqnarray}
Q^1 D \bar{Q}^{\dot{1}}&=&\Delta^{-2}\int d\omega d \omega' \sum \vert Y \vert^{2} e^{-i(k-k')r*} (S_-S_-') e^{i(m-m')\phi} e^{-i(\omega- \omega') t} \nonumber \\
& & \times [-i(r^2 + a^2)(w+k) - (r-M) + ima]
\end{eqnarray}

\begin{eqnarray}
\bar{Q}^{\dot{1}} D Q^1&=&\Delta^{-2}\int d\omega d \omega' \sum \vert Y \vert^{2} e^{-i(k-k')r*} (S_-S_-') e^{i(m-m')\phi} e^{-i(\omega- \omega') t} \nonumber \\
& & \times [i(r^2 + a^2)(w'+k') - (r-M) - im'a] \label{term4}
\end{eqnarray}

where the summation is over $l,m,l',m'$, $\vert Y \vert^{2}=Y_{\rm{hole}} Y_{\rm{hole}}^*$ and $S_{\pm}=[{}_{\pm 1/2}S_{lm}(\theta , a\omega)]$.
Note that all of the four terms  (\ref{term1}-\ref{term4}) that construct $T_{ab}l^al^b$ in (\ref{Tablalb}) have a common factor of $\Delta^{-2}$ which will cancel with the $\Delta^2$ term in the first term of the left-hand-side of (\ref{lalbnanb0}). We proceed to evaluate the terms in (\ref{Tabnanb})

\begin{eqnarray}
P^0 D \bar{P}^{\dot{0}}&=& (\Delta / 2 \Sigma )\int d\omega d \omega' \sum \vert \rho \vert^{2}  \vert Z \vert^{2}e^{-i(k-k')r*} (S_-S_-') e^{i(m-m')\phi} e^{-i(\omega- \omega') t} \nonumber \\
& & \times [i(r^2 + a^2)(w'-k')- \Delta \rho^* - (r-M) - im'a]
\end{eqnarray}

$P^0 D \bar{P}^{\dot{0}}$ and the remaining three terms in $T_{ab}n^an^b$ terms turn out to be at least first order in $\Delta$ so the second term on the left-hand-side of (\ref{lalbnanb0})  will not contribute to the flux at a surface at the horizon where $\Delta \to 0$. 

Now we can evaluate (\ref{eq:dC/dt1}) and (\ref{deltaccc}) for a surface at the horizon $(r=r_+ , \Delta \to 0 \Rightarrow M^2+a^2=2Mr_+ )$ of an extremal Kerr black hole $(r_+=M)$. Note that $\sqrt{-g}=\Sigma \sin \theta $ at the horizon. We first take the time integral in (\ref{deltaccc}) which gives a delta function in $\omega$ and $\omega'$ allowing us to evaluate the integral over $\omega'$. Now the angular functions are all functions of $\omega$ so we can use the orthonormality relation (\ref{anguortho}) and evaluate $\theta$ and $\phi$ integrals. This gives the kronecker delta $\delta_{ll'}\delta_{mm'}$ so we evaluate the sums over $l'$ and $m'$. Having $\omega=\omega'$ (from the integration of the delta function in $\omega$ and $\omega'$ over $d\omega'$) and $m=m'$ (as a result of $\theta$ and $\phi$ integrals using the orthonormality relation) leads to $k=k'$. The expressions in the square brackets in (\ref{term1}-\ref{term4}) reduce to $\pm i2Mr_+(k + \omega -am/(2Mr_+))=\pm i2Mr_+(2k)$. Finally  $\delta ({\mathcal{C}})$ takes the form
\begin{equation}
\delta {\mathcal{C}}  =   \int  d\omega  \sum_{lm}\vert Y \vert^{2} 8r_+k \label{deltac}
\end{equation}
The expression (\ref{deltac}) becomes negative in the region $0<\omega <am/(2Mr_+)$ where $k$ is negative. Hence a wave packet with dominant frequency in this region will lead to violation of cosmic censorship.

The violation of CCC in this gedanken experiment is essentially different from similar attempts to overspin a Kerr black hole \cite{Jacobson-Sot,overspin} involving test bodies and massless bosonic test fields, respectively. These experiments consider nearly extremal Kerr black holes parametrized by a dimensionless quantity $\epsilon \ll 1$ such that $J/M^2=a/M=1-2\epsilon^2$. In both cases $\delta J$ has a lower limit so that $J+\delta J>(M+\delta E)^2$ and CCC is violated. In the case of test bodies an upper limit is brought to $\delta J$ by requiring that the test body crosses the horizon; whereas in the case of bosonic test fields there exists a lower limit for the frequency $\omega$ to avoid superradiance, which constitutes an upper limit for $\delta J$. In both experiments the allowed range of $\delta J$ and $\delta E$ to violate CCC is of order $\epsilon^2$ and $\delta {\mathcal{C}} \sim -\epsilon^2$. (This range and $\delta {\mathcal{C}}$ vanishes as $\epsilon \to 0$ so extremal black holes cannot be overspun.) Later dissipative (radiative) and conservative (gravitational) self-force effects were considered for \cite{Jacobson-Sot} (the case of test bodies) and it was shown that conservative self-force is comparable to the terms giving rise to naked singularities~\cite{barrauseEtal1}. In principle, gravitational and radiative self-force effects can also be calculated for the case of massless fields to compensate for negative $\delta {\mathcal{C}}$. However, for neutrino fields $\omega$ can be lowered to zero so that  $\delta J$ and the magnitude of $\delta {\mathcal{C}}$  grows without bound. Self-force effects may at most compensate for negative  $\delta {\mathcal{C}}$ in a small region of frequencies near the superradiant limit, of width $\sim \epsilon^2$. Therefore self-force effects cannot prevent the horizon from being destroyed in the classical treatment of neutrino fields.
 
Another back reaction of interest is the possibility of formation of another horizon outside the original black hole as it is destroyed. In \cite{hod3} a charged shell (mass $m$, charge $q$) is adiabatically lowered to a Reissner-Nordstr\"{o}m black hole ($M$, $Q$). By Birkhoff theorem the space-times inside and outside the shell are described by R-N metric with parameters $[M,Q]$ and $[(M+E(r)),(Q+q)]$, respectively. ($E(r)$ is the total energy of the shell consisting of its rest mass red shifted by the gravitational field, its electrostatic interaction with the black hole, and electrostatic and gravitational self-energy.) The equation for the horizon radius outside the shell is $1-2(M+E(r))/r + (Q+q)^2/r^2=0 $, and it has a real solution outside the original black hole, thus a new horizon is formed. This is not directly applicable to the case of massless fields and Kerr black holes. Let us localize the fields as particles with energy $\hbar \omega$ and angular momentum $ \hbar m$. The equation for a horizon radius outside a sphere enclosing the original black hole and the particles is $r^2 +2(M+\delta(M))r + (a+\delta(a))^2=0$. The energy and angular momentum of the field (particles) is proportional to $\omega$ and $m$ and do not explicitly depend on $r$. The equation for a horizon radius is still quadratic with constant coefficients and has no real roots when  $ (M+\delta(M))^2< (a+\delta(a))^2$, that is $\omega< (am)/(2Mr_+)$. The process leading to the formation of a shielding horizon does not apply to the case of massless fields interacting with Kerr black holes.

\section{Conclusions}
We showed that the classical interaction of an extremal Kerr black hole with a test neutrino field can destroy the horizon of the black hole and convert it to a naked singularity. We also briefly discussed the backreaction due to radiative and gravitational self-force effects and argued that the destruction of black holes by neutrino fields cannot be fixed by these effects. A hope to avoid the naked singularity lies in the quantum treatment of the problem. Unruh studied the second quantization of neutrino fields in Kerr background  and found that black holes spontaneously emit neutrino fields  in superradiant modes and spin themselves down \cite{unruh}. This outward flux equals the amount later calculated by Hawking in the limit that the  temperature $\kappa/2\pi$ (where $\kappa$ is the surface gravity) is low \cite{hawkingrad}. 
In that work Unruh refers to a possibility suggested by Feynman: if we send in neutrino fields with frequency $\omega<am/(2Mr_+)$ towards the black hole, part of this flux may be suppressed because of exclusion principle. This represents a quantum form of superradiance. To the extent that we accept that quantum emission of radiation  fills the phase space, one may at best be able to suppress the constant spinning down of the black hole. In that case the modes with $0<\omega<am/(2Mr_+)$ would be harmless because they are either not absorbed or a field in the same mode has been emitted so they do not contribute to the flux at the horizon. 

The region in which the expression (\ref{deltac}) becomes negative is exactly the superradiant region for bosonic fields \cite{misner,teuknature}. If the frequency of the incoming neutrino field is in this region, the interaction of the field with the black hole leads to the destruction of the horizon, exposing the singularity of the black hole to outside observers. However in almost identical analysis involving bosonic fields and extremal black holes (see \cite{dkn} and \cite{duztasem} for scalar and electromagnetic cases), cosmic censorship conjecture has been shown to remain valid (the horizons of extremal black holes can not be destroyed using integer spin test fields). This is in accord with the fact fermionic fields do not exhibit superradiant behaviour unlike bosonic fields  \cite{chandrabook}. With the possible proviso about quantum effects discussed in the previous paragraph, there can be a net absorption of the superradiant modes which carry more angular momentum than energy. The destruction of black holes by neutrino fields  cannot be fixed by self-force effects unlike the cases involving test bodies and bosonic fields. Hence we conclude that the classical interaction of an extremal Kerr black hole with a test neutrino field can lead to a generic violation of the cosmic censorship conjecture.

\section*{Acknowledgements}
I would like to thank {\. I}.  Semiz for helpful discussions and comments.


\section*{References}
\begin{thebibliography}{99}
\bibitem{singtheo} Hawking S W and  Penrose R 1970 The Singularities of Gravitational Collapse and Cosmology {\it Proc. R. Soc. London}  \textcolor{blue}{ \textbf{314} 529-48 }


\bibitem{waldbook}  Wald R M  1984 {\it General Relativity} (London: The University of Chicago Press )

\bibitem{tipler}  Tipler  F J 1976 Causality Violation in Asymptotically Flat Space-Times {\it  Phys. Rev. Lett.} \textcolor{blue}{ \textbf{37} 879-82}

\bibitem{penrose.orig.ccc} Penrose R 1969 Gravitational Collapse : The Role of General Relativity {\it Riv. Nuovo Cimento} \textcolor{blue}{ \textbf{1} 252-76 }

\bibitem{waldrev} Wald R M 1997 Gravitational Collapse and Cosmic Censorship arXiv:gr-qc/9710068 


\bibitem{wald74} Wald R M 1974 Gedanken Experiments to Destroy a Black Hole {\it Ann. Phys.} \textcolor{blue} {\textbf{82} 548-56}

\bibitem{ri_saa_1}  Richartz M and Saa A 2008 Overspinning a nearly extreme black hole and the Weak Cosmic Censorship conjecture {\it  Phys. Rev. D} \textcolor{blue} { \textbf{78} 081503}

\bibitem{ri_saa_2}  Richartz M and Saa A 2011 Challenging the weak cosmic censorship conjecture with charged quantum particles {\it  Phys. Rev. D} \textcolor{blue} { \textbf{84} 104021}

\bibitem{dkn}  Semiz \.{I}. 2010 Dyonic  Kerr-Newman black holes, complex scalar field and Cosmic Censorship {\it  Gen. Relativ. Gravit.} \textcolor{blue}{\textbf{43} 833-46}

\bibitem{duztasem} D\"{u}zta\c{s} K 2014 Electromagnetic field and cosmic censorship { \it Gen. Relativ. Gravit.} \textcolor{blue}{\textbf{46} 1709 }


\bibitem{newpen} Newman E and Penrose R 1962 An Approach to Gravitational Radiation by a Method of Spin Coefficients  {\it  J. Math. Phys.} \textcolor{blue}{ \textbf{3} 566-78}

\bibitem{chandradirac} Chandrasekhar S 1976 The solution of Dirac's equation in Kerr geometry { \it Proc. R. Soc. Lond. A} \textcolor{blue}{ \textbf{349} 571-75}

\bibitem{teuk2} Teukolsky S A 1973 Perturbations of a Rotating Black Hole. I. Fundamental Equations for Gravitational, Electromagnetic, and Neutrino-Field Perturbations {\it  Astrophys. J.}  \textcolor{blue}{ \textbf{185} 635-47}

\bibitem{agr} Stewart J 1991  {\it Advanced General Relativity} (Cambridge: Cambridge University Press) 


\bibitem{teukII} Teukolsky S A and  Press W H 1973  Perturbations of a Rotating Black Hole. II. Dynamical Stability  of the Kerr metric  { \it Astrophys. J.} \textcolor{blue}{\textbf{185}  649-73}


\bibitem{teukIII}  Teukolsky S A and  Press W H 1974 Perturbations of a Rotating Black Hole. III. Interaction of the Hole with Gravitational and Electromagnetic Radiation {\it  Astrophys. J. } \textcolor{blue}{\textbf{193} 443-61}

\bibitem{guven} G\"{u}ven R 1977  Wave Mechanics of Electrons in Kerr Geometry {\it  Phys. Rev. D } \textcolor{blue}{ \textbf{16} 1706-11}

\bibitem{scatter}  Futterman J A H  Handler F A and Matzner R A 2009 {\it Scattering from black holes} (Cambridge: Cambridge University Press) 


\bibitem{Jacobson-Sot} Jacobson T and  Sotiriou T P 2009  Over-spinning a black hole with a test body {\it Phys. Rev. Lett.} \textcolor{blue}{\textbf{103} 141101}

\bibitem{overspin}  D\"{u}zta\c{s} K and   Semiz \.{I} 2013 Cosmic censorship, black holes and integer-spin test fields {\it Phys. Rev. D }\textcolor{blue}{\textbf{88} 064043} 

\bibitem{barrauseEtal1} Barausse E Cardoso V and Khanna G 2010 Test Bodies and Naked Singularities: Is the Self-Force the Cosmic Censor? {\it Phys. Rev. Lett.} \textcolor{blue}{\textbf{105} 261102}

\bibitem{hod3} Hod S 2013 Cosmic censorship: Formation of a shielding horizon around a fragile horizon {\it Phys. Rev. D} \textcolor{blue}{\textbf{67} 024037 }  

\bibitem{unruh}  Unruh W G 1974 Second quantization in the Kerr metric {\it Phys. Rev. D} \textcolor{blue}{\textbf{10} 3194-3204 }

\bibitem{hawkingrad}  Hawking S W 1975 Particle creation by black holes {\it Commun. Math. Phys.} \textcolor{blue}{\textbf{43}  199-220}

\bibitem{misner}  Misner C W 1972 Interpretation of Gravitational-Wave Observations   {\it Phys. Rev. Lett.} \textcolor{blue}{ \textbf{28} 994-97}

\bibitem{teuknature}Press W H and   Teukolsky S A 1972 Floating Orbits, Superradiant Scattering and the Black-hole Bomb { \it Nature} \textcolor{blue}{\textbf{238} 211-12 }

\bibitem{chandrabook} S. Chandrasekhar 1983 {\it The Mathematical Theory of Black Holes} (New York: Oxford University Press)


\end{thebibliography}
\end{document}